\RequirePackage{fix-cm}
\documentclass[smallextended]{svjour3}       
\smartqed  
\usepackage{graphicx}
\usepackage{amssymb}
\usepackage{color}
\usepackage{graphicx}
\usepackage{amsmath}
\usepackage{subfigure}
\usepackage{rotating}
\usepackage{multirow}
\usepackage{array}
\usepackage{algorithm}
\usepackage{algorithmic}

\journalname{annales des telecommunications}
\begin{document}

\title{Investigating Quality Routing Link Metrics in Wireless Multi-hop Networks}

\author{N. Javaid, A. BiBi, A. Javaid, Z. A. Khan, K. Latif,  M. Ishfaq}


\institute{Dr. Nadeem Javaid, Miss. Ayesha BiBi, and Mr. Kamran Latif are with the Center for Advanced Studies in Telecommunications (CAST), COMSATS Institute of Information Technology, Islamabad, Pakistan. \\Dr. Akmal Javaid is with the Department of Mathematics, COMSATS Institute of Information Technology, Wah Cant, Pakistan. \\Dr. Zahoor Ali Khan is with the Internetworking Program, Faculty of Engineering, Dalhousie University, Halifax, Nova Scotia, Canada. \\Prof. Dr. Mohammad Ishfaq is with King Abdulaziz University, Rabigh, Saudi Arabia. \\
Corresponding author's Email: nadeemjavaid@comsats.edu.pk, nadeemjavaid@yahoo.com, Web: www.njavaid.com
   }

\date{Received: date / Accepted: date}

\maketitle

\begin{abstract}
In this paper, we propose a new Quality Link Metric (QLM), ``Inverse Expected Transmission Count (InvETX)'' in Optimized Link State Routing (OLSR) protocol. Then we compare performance of three existing QLMs which are based on loss probability measurements; Expected Transmission Count (ETX), Minimum Delay (MD), Minimum Loss (ML) in Static Wireless Multi-hop Networks (SWMhNs). A novel contribution of this paper is enhancement in conventional OLSR to achieve high efficiency in terms of optimized routing load and routing latency. For this purpose, first we present a mathematical framework, and then to validate this frame work, we select three performance parameters to simulate default and enhanced versions of OLSR. Three chosen performance parameters are; throughput, Normalized Routing Load and End-to-End Delay. From simulation results, we conclude that adjusting the frequencies of topological information exchange results in high efficiency.
\end{abstract}

\section{Introduction}
Communication at any time without any disruption for mobile users is provided by Wireless Multi-hop Networks (WMhNs). These networks have some distinguished features due to dynamic topologies; various number of nodes, communication demand at random times, in random directions and for different sessions. Underlying network demands a routing protocol to dynamically cope with changing topologies. Mobile nodes in WMhNs are very often limited in resources such as processing capabilities, storage capacity, battery power, bandwidth etc. This implies that the routing protocol must be able to minimize the control traffic, (as trigger/periodic update messages), delays (due to retransmissions, or computation of metrics), and so on. The performance of a wireless network depends upon efficiency of the routing protocol operating it. The most important component of the routing protocol is `routing link metric'. Because, a link metric first considers the quality routes then decides the best end-to-end path. A Quality Link Metric (QLM) plays a key role to achieve the desired performance from an underlying network by making the routing protocol: fast enough to adopt topological changes, light-weight to minimally use the resources of nodes, intelligent to select the fastest path from source to destination among available paths and capable to enable the nodes to have a comprehensive idea about the topology.

Considering demands of the underlying network from its operating protocol and factors influencing its performance, a QLM is supposed to fulfill certain requirements. An efficiently designed QLM better helps a routing protocol to achieve appreciable performance by dealing with these issues.
In our previous work [1], design requirements of QLM are discussed in detail. Moreover, a new QLM; InvETX, is also compared with Expected Transmission Count (ETX) [2], Minimum Loss (ML) and Minimum Delay (MD) [4] in Optimized Link State Routing (OLSR) [5] protocol. As, we discuss earlier that a routing protocol and link metric collectively are responsible for efficient performance, therefore, we address improvement in routing protocols as well. For this purpose, in this paper, we enhance OLSR to reduce routing overhead. To validate our enhancements, we evaluate and compare the performance of selected metrics in default and enhanced OLSR.

Among a wide range of reactive and proactive protocols, we have selected the proactive class, because: (i). proactive routing protocols are designed for static networks, so, they perform excellent in such networks [6]. We therefore, have chosen a proactive protocol, (ii). OLSR is designed for static and dense networks [7], (iii). in [8], we have carried out almost the same study in the same scenario with Destination-Sequenced Distance Vector (DSDV) [9]. So, to provide the readers with another proactive protocol, we have chosen OLSR.

\section{Related Work and Motivation}
After analyzing reactive and proactive protocols, Yang \textit{et al.} [10] deduce that proactive protocols that implement the hop-by-hop routing technique, as DSDV and OLSR protocols are the best choice for Static Wireless Mesh Networks (SWMNs). They also inspect design requirements for routing link metrics for the mesh networks and related them to the routing techniques and routing protocols.

Das \textit{et al.} in [11], discuss dynamics of the well known metrics: ETX, Expected Transmission Time (ETT) [12] and Link Bandwidth [13] in real test beds. Across various hardware platforms and changing network environments, they test two requirements: stability and sensitivity for some existing routing link metrics. Authors also discuss the dynamics of the above mentioned metrics and tested their performance on the test beds for the above stated requirements.
	
In [14], Yaling \textit{et al.} systematically analyze impact of working of wireless routing link metrics on the performance of routing protocols. Three operational requirements: optimality, consistency and loop-freeness is also discussed. However, these properties do not cover all design requirements; for example, computational overhead, a metric can produce and the performance trade-offs a metric has to make among different network performance factors. For example, a routing protocol achieves higher throughput at the cost of end-to-end delay or routing overhead. Therefore, we analyze possible design requirements in [1]. Further, in our previous work, we implement three existing link metrics; ETX, MD,and ML, and one newly proposed metric; InvETX OLSR using NS-2.34. The simulation results show that how computational burden of a metric degrades performance of the respective protocol and how a metric has to trade-off between different performance parameters.

In fact this work is solely devoted to the ``Static Wireless Networks''. To keep focused, we provide the readers with an in-depth analysis from Mathematical Framework to Simulations for ``Static Wireless Networks'' only. In our other works, we have extensively analyzed Mobility in ``Dynamic Wireless Networks'' (with varying mobilities and speeds) and have compared both proactive and reactive routing protocols. In the text below, we are providing our works in which we have already analyzed, ``dynamic wireless mesh networks''.

Being an interface between the underlying wireless network and mobile users, a routing protocol plays an important role. So, to provide the reader with a comprehensive idea about routing and how do the routing protocols react to the topological changes [6], we have chosen the most widely experimented and frequently used protocols for our study; three from reactive or on-demand class: AODV, DSR, DYMO, and three from proactive or table-driven class DSDV, FSR, OLSR.

In[15], we have modeled the routing overhead generated by three reactive routing protocols; AODV, DSR and DYMO. We have modeled the energy consumed and time spent per packet both for route discovery and route maintenance. The proposed framework is evaluated in NS-2 to compare performance of the chosen routing protocols.

[16] evaluates and compares the performance of two routing protocols, one is reactive; DYMO and other is proactive; OLSR in Mobile Ad-hoc Networks and Vehicular Ad-hoc Networks. Performance of these protocols is analyzed using three performance metrics; Packet Delivery Ratio, Normalized Routing Overhead and End-to-End Delay against varying scalabilities of nodes. We perform these simulations with NS-2 using TwoRayGround propagation model.

In [17], we simulate the three routing protocols; DSDV, OLSR, DYMO in NS-2 to evaluate and compare their performance using two Mac-layer protocols 802.11 and 802.11p. Comprehensive stimulation work is done for each routing protocol. With the performance metrics; Throughput, End to End Delay (E2ED), Normalized Routing load (NRL) all of the three protocols are evaluated, analyze and compared in the scenarios of varying the mobilities and scalabilities.

[18] evaluates, analyzes, and compares the impact of mobility on the behavior of three reactive protocols (AODV, DSR, DYMO) and three proactive protocols (DSDV, FSR, OLSR) in multi-hop wireless networks. We take into account throughput, end-to-end delay, and normalized routing load as performance parameters. Based upon the extensive simulation results in NS-2, we rank all of six protocols according to the performance parameters.

In this paper, we present mathematical framework to analyze requirements of routing protocol as well as link metric to achieve efficient performance in underlying wireless network. To improve overall efficiency in SWMhNs, we enhance OLSR; EOLSR, and than evaluate performance of default OLSR; OLSR, with EOLSR. To check the effect of enhancement over link metrics, we compare OLSR and EOLSR with ETX, ML, MD and InvETX. To check the efficiency of enhancement in terms of successfully delivered data, routing load and routing latencies we select three performance parameters for analytical comparison: throughput, End-to-End Delay (E2ED) and Normalized Routing Load (NRL), respectively.

\begin{table}[H]
\caption {Design Requirements of Routing Link Metric}
\begin {center}
 \begin{tabular}{|p{2cm}|p{4cm}|p{5cm}|p{2cm}|}
\hline
\textbf{Design Requirement}& \textbf{Issue} & \textbf{Possible Solution} & \textbf{Metric or Algorithm}\\\hline
Minimizing path length & Longer path increases routing latency and reduces throughput of a path & 1)By minimizing number of transmissions and, 2) Paths selection with minimum loss rates or higher probabilities of successful transmissions, etc. & Hop count\\\hline

Balancing traffic load & Overloaded traffic causes drop rate due to congestion & Divert traffic from congested path or overloaded nodes to underloaded or idle ones & Transmissions reduction\\\hline

Minimizing delay & Delay results time-out-buffer & Delay can be reduced with selection of a path having minimum intra-flow and inter-flow interferences along with queuing delays, and maximum link capacity & ETX, Per hop RTT, Per hop PktPair, ML\\\hline

Maximizing aggregating bandwidth & network's capacity directly effects throughput & 1) Minimize interferences or retransmissions, and 2) allowing the multiple rates to coexist in a network where a higher channel rate is used over each link & MIC\\\hline

Minimizing energy consumption & A path with an unreliable link produce longer delay due to higher retransmission rates and ultimately results in raise in energy consumption & Reduction in retransmissions during routing to optimize communication delay & MTPR, MBCR\\\hline

Minimizing channel/interface switching & Data flows switching on different channels results in delay & Interface assignment strategy keeps one interface fixed on a specific channel, while other interfaces can be switched among the remaining channels, when necessary & MIC, WCETT \\\hline

Minimizing the Computational Overhead & Computational overhead consumes memory, processing capability and battery power & computations should be considered that must not consume memory, processing capability and the most important; battery power & InvETX\\\hline

Minimizing interference & Intra-flow and inter-flow interferences result in bandwidth starvation & During path calculation capture diversity of channel assignments and link capacity & MCR Protocol\\\hline

Maximizing route stability & Instability in path's weight results in drop rates &  Load sensitivity or topology-dependent metrics solve instability issues & Link affinity metric with MCMR\\\hline

Maximizing fault tolerance/minimizing route sensitivity & Faulty routes cause drop rates in high network flows & This problem can be solved through providing redundant information of alternative paths & wireless ad-hoc networks\\\hline

Avoiding short and long lived loops & Redundant links due to short and long lived loops results in more path lengths and consequently increased E2ED & 1) minimum TTL value that eliminates mini-loops Faheem \textit{et al.} 2) Fresh sequence number etc. & OLSR in Sparse WMN's \\\hline

Considering performance trade-offs & E2ED in static networks cause drop rate & A suitable trade-off helps to increase efficiency & ML \\\hline
\end{tabular}
\end{center}
\end{table}

\section{Mathematical Framework}
The factors affecting the wireless networks help to have an idea about the problems they have to face. Along with other protocols that operate a
network, routing protocols play a significant role in the performance of WMhNs. In WMhNs, specially in SWMhNs, generally the link quality considerably varies in different periods of time. The reasons may be; some mobile nodes are moving randomly, some go-out of range, some intentionally cut-off the ongoing communication, some die-out due to battery and so on. The respective routing protocols must be able to dynamically cope with the situation.

As, this work is devoted for routing protocol and link metric behavior in SWMhNS, therefor, a linear programming model is first constructed to analyze performance efficiency. In this model, effect of capacity of a link along with nonzero constraints are listed for path selection requirements in a link metric. Routing load and routing latency of proactive protocols, $pro$, are modeled in [19]. To address link metric efficiency in routing protocols, we first construct an integer linear programming model to list possible constraints against routing overhead. Let $e$ denotes efficiency and is considered as objective function as:

\begin{eqnarray}
max\,\,\,e
\end{eqnarray}

\textbf{Subject to:}
\begin{equation*}
\begin{aligned}
\begin{split}
& \sum_{j:(i,j)\in E}x_{ij} - \sum_{j:(j,i)\in E}x_{ji} = \begin{cases} \displaystyle e\,\,\,\,\, i=s, \\
                                                                        \displaystyle 0\,\,\,\,\, i\in N \backslash \{s,t\},\text{(1.a)}\\
                                                                        -e\,\,\,\,\, i=t.  \,\,\,
                                                                       \end{cases}\\
&0\le x_{ij} \le Cap_{ij}.z_{ij}, (i,j)\in E\,\,\,\,\text{(1.b)}\\
&\sum_{j:(i,j)\in E} Z_{ij}, (i,j)\in {0,1}\,\,\,\,\text{(1.c)}\\
&\sum_{j:(i,j)\in E} C_{E-per}^{pro}+C_{E-tri}^{pro}+C_{E-metric}^{QLM} \begin{cases} \displaystyle =\beta_{cri}\,\,\,\,\, e=0, \\
                                                                        \displaystyle < \beta_{cri} \,\,\,\,\, e\neq 0. \,\,\,\,\text{(1.d)}\\
                                                                       \end{cases}\\
&\sum_{j:(i,j)\in E} C_{T-per}^{pro}+C_{T-tri}^{pro}+C_{T-metric}^{QLM}\begin{cases} \displaystyle =\tau_{cri}\,\,\,\,\, e=0, \\
                                                                        \displaystyle < \tau_{cri} \,\,\,\,\, e\neq 0. \,\,\,\,\text{(1.e)}\\
                                                                       \end{cases}\\
\end{split}
\end{aligned}
\end{equation*}
\normalsize \newline where, $Cap_{ij}$ specifies maximum achievable link rate over link  $L_{ij}.z_{ij}=1$ in such a way that $L_{ij}$ may have a nonzero flow. Eq. 1(c) denotes that there is at most one outgoing link from each node with a nonzero flow. Eqs. 1(a)(b)(c) are subjective constraints for single route. The links along that path have the same flow and all other links have zero flow. As, we address the issues regarding routing protocol along with routing link metric.  Therefore, in Eq. 1(d)(e), we model routing overhead of a routing protocol in terms of routing load and routing latency. $C_{E-per}^{pro}$ and $C_{E-tri}^{pro}$ denote cost of energy consumed per packet by a $pro$ for periodic and trigger updates, respectively. Whereas, $C_{T-per}^{pro}$ and $C_{T-tri}^{pro}$ in Eq. 1(e) are routing latencies of periodic and trigger periods, respectively. For a QLM, $C_{E-metric}^{QLM}$ and $C_{T-metric}^{QLM}$ are routing load and routing latencies, respectively. $\beta_{cri}$ and $\tau_{cri}$ denote critical bandwidth and critical value of time.

Usually, behavior of the channels varies in links and then in complete paths from source to destination. Thus, effecting capacity of link in a network is an  obvious constraint in eq. 1(a), along with nonzero flow constraints (eq. 1(b)(c)). In the case of Quality of Service (QoS) routing, the the link creating bottle neck for performance must be given attention. Similarly, change in the quality of one link affects the others. Design requirements for routing algorithms and link metrics are summarized in Table. 1.

\subsection{Framework of Routing Overhead in OLSR}
In this work, we are dealing with SWMhNs which are bandwidth sensitive because of limited bandwidth. One of the reasons for bandwidth consumption is routing overhead generated by a routing protocol. For efficient utilization of resources in SWMhNs, we select OLSR because it reduces redundant retransmissions more efficiently in static networks due to Multi-Point Relay (MPR) scheme. OLSR functionality with MPR selection scheme is presented in next sub section. In our work we have studied OLSR for degree based routing approach. OLSR use HELLO and Topology Control (TC) messages during route computation. Let $C_{E-total}^{OLSR}$ denotes total cost of energy consumed per packet by OLSR and is the sum of  $C_{E-HELLO}^{OLSR}$ and $C_{T-TC}^{OLSR}$.

\begin{figure*}[!t]
  \centering
 \subfigure[Simple Network]{\includegraphics[height=6 cm,width=7 cm]{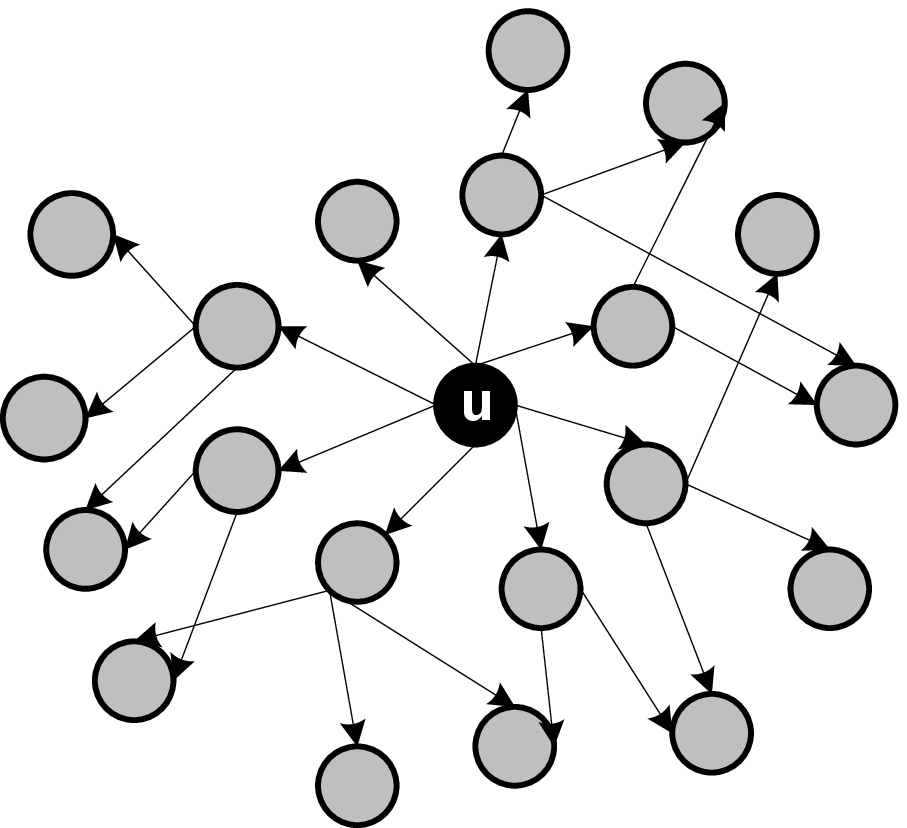}}
 \subfigure[MPRs Selection]{\includegraphics[height=6 cm,width=7 cm]{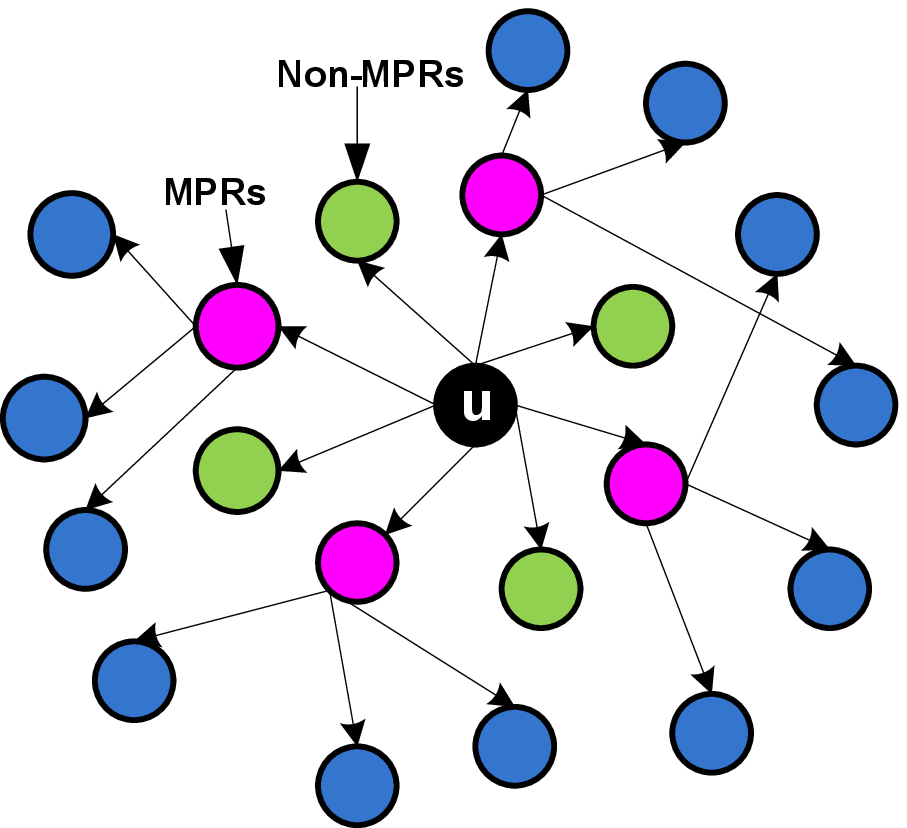}}
  \caption{Figure.1. MPRs mechanism in OLSR}
\end{figure*}

\begin{eqnarray}
C_{E-total}^{OLSR} =C_{E-HELLO}^{OLSR} + C_{E-TC}^{OLSR}
\end{eqnarray}

OLSR disseminates only trigger updates for maintaining fresh routes. The interval for transmission of routing updates varies with respect to the status of MPR. TC messages are transmitted through default interval, if MPR status remains the same. OLSR triggers routing updates whenever, change in the status of MPR nodes occurs.

Let trigger cost of TC messages due to MPR redundancy is $C_{E-TC-trig}$ and default cost of TC messages due to the stable MPRs is $C_{E-TC-def}$. Whereas, $\tau_{hello}$ specifies HELLO interval and $\tau_{NL}$ is the total network life time. Moreover, we have defined three sets of nodes; \textit{(i)} connected neighbor nodes; $N_{br}$, \textit{(ii)} selected MPRs $(MPRs)$ and \textit{(iii)} all nodes in the network $N$. So, $C_{E-total}^{OLSR}$ in eq. 2 can be written as:

\begin{equation}
C_{E-total}^{OLSR}=C_{E-HELLO}+C_{E-TC-trig}+C_{E-TC-def}
\end{equation}\newline where,

\begin{equation}
C_{E-HELLO}=\frac{\tau_{NL}}{\tau_{hello}}\sum _{ \forall i \in N} \sum_{\forall j \in Nbr}j
\end{equation}

\begin{equation}
C_{E-TC-trig}=\int^{\tau_{NL}}\sum_{\forall i\in N}\sum_{\forall j\in MPRs}|Sgn(change_j^{MPR})|j
\end{equation}

\begin{equation}
C_{E-TC-def}=\int^{\tau_{NL}}\sum_{\forall i\in N}\sum_{\forall j\in Nbr}|Sgn(change_j^{MPR})|j
\end{equation}

The trigger updates of OLSR depend upon $change_j^{MPR}$. MPR selection is based on maximum degree selection to solve NP-complete problem. A framework for MPR selection is presented as follows:

\subsubsection{Framework of MPR Selection}
Using graph theory, a wireless network can be defined as a bidirectional undirected graph $G(V,E)$, such that $V=$ vertices. Nodes $m$ and $n$ share a bidirectional link $(m, n)$. For bidirectional links, if and only if nodes $m$ and $n$ hear each other, they can communicate. Let $H_{1}(u)$ be the 1-hop neighbors of node $u$. Let $H_{2}(u)$ denotes 2-hop neighbors of $u$ (neighbors of 1-hop neighbors of $u$ but not 1-hop neighbors of $u$).

Let, $d_{max}$ denotes the maximum degree of a node in the graph.

\begin{eqnarray}
d_{max} = max_{u,V} H_{1}(u)
\end{eqnarray}

MPR selection is based on willingness in 1-hop neighborhood $H_{1}(u)$. The reachability of a node with second hop $H_{2}(u)=v$ neighbors $d_{u}^{+} (v)$ is defined as:

\begin{eqnarray}
d_{u}^{+} (v) =\mid {w \in H_{1}(v) \mid v\in H_{1}(u)\,\,and\,\,w\in H_{2}^{+} (u) }
\end{eqnarray}

The maximum reachability of a node belonging to $H_{1}(u)$),$d_{u}^{+}$, is presented as:

\begin{eqnarray}
d_{u}^{+} = max_{n\in N(u)} d_{u}^{+}(v)
\end{eqnarray}

Also, the $MPR(u)$ set of a node $(u)$, a subset of $H_{1}(u)$, can be defined as follows:

\begin{eqnarray}
\forall w\in H_2 (u), \exists v\in MPR(u) \,\,\,\,\,\,\,\,such\,\,that\,\,w\in H_{1}(v)
\end{eqnarray}

\subsection{Framework of Selected Link Metrics}
While designing a routing metric, necessary computations should be considered that must not consume memory, processing capability and the most important; battery power. For example, we discuss the case of three widely used QLMs for wireless routing protocols: \textit{ETX}, its  inverse, say, \textit{InvETX} and \textit{ML}.

For an end-to-end path, $P$, these metrics are  expressed by following equations:

\begin{eqnarray}
 ETX_{P}=\sum_{l \in P}^{}\frac{1}{fd(l)\times rd(l)}
\end{eqnarray}

\begin{eqnarray}
 InvETX_{P}=\sum_{l \in P}^{}{fd(l)\times rd(l)}
\end{eqnarray}

\begin{eqnarray}
  ML_{P}=\prod_{l \in P}^{}fd(l)\times rd(l)
\end{eqnarray}

\begin{figure}[h]
  \centering
 \subfigure{\includegraphics[height=6 cm,width=7 cm]{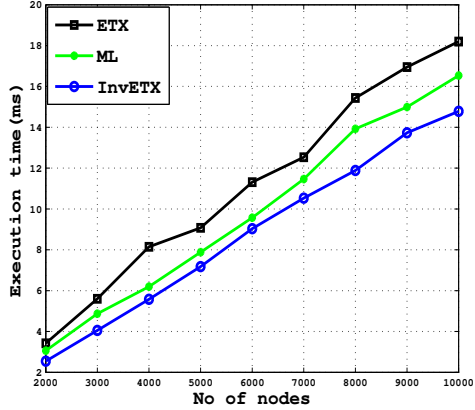}}
  \caption{Computational Overhead by ETX, ML and InvETX}
\end{figure}

Where, $fd(l)\times rd(l)$ is the probability of success for delivery of probe packets (134 $bytes$ each) on the link $l$ on $P$ from source to destination (forward direction) and from destination to source (reverse direction).

Regarding computational complexity, all of the three metrics have to calculate the equal number of products $fd(l) \times rd(l)$ for the same number of links. But $ETX$ has to suffer from more computational overhead (inverse and sum of $n$ products) than $ML$ (multiplication of $n$ products only). Similarly, $ML$ generates more computational overhead than $InvETX$. As a result, $InvETX$ achieves higher throughputs than $ML$ and $ETX$. Similarly, $ML$ performs better than $ETX$.

\begin{figure}[ht]
  \centering
 \subfigure[Computational Overhead of ETX, ML, InvEX]{\includegraphics[height=6 cm,width=7 cm]{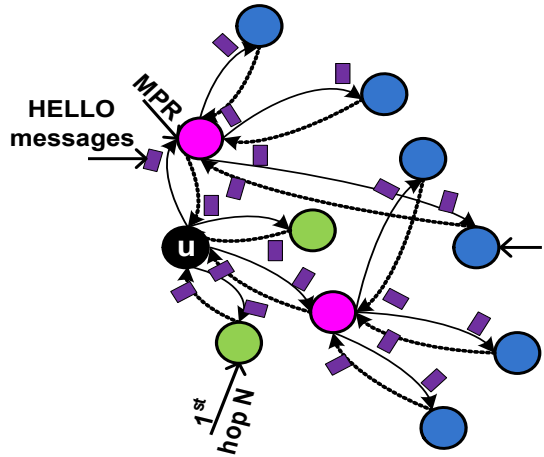}}
 \subfigure[Computational Overhead of MD]{\includegraphics[height=6 cm,width=7 cm]{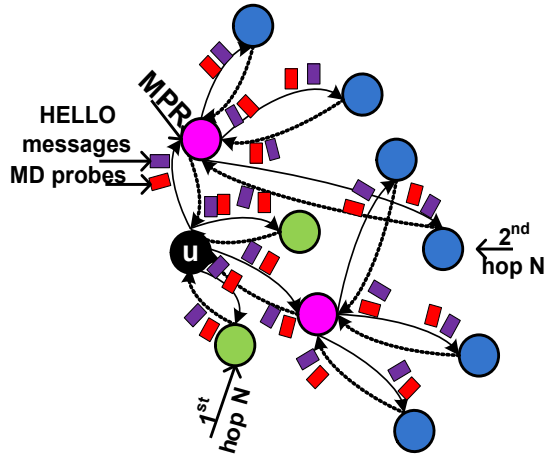}}
  \caption{Figure.2. Link Metric Selection}
\end{figure}

\begin{figure}[ht]
  \centering
 \subfigure[Throughput of OLSR with 4 metrics]{\includegraphics[height=4 cm,width=5 cm]{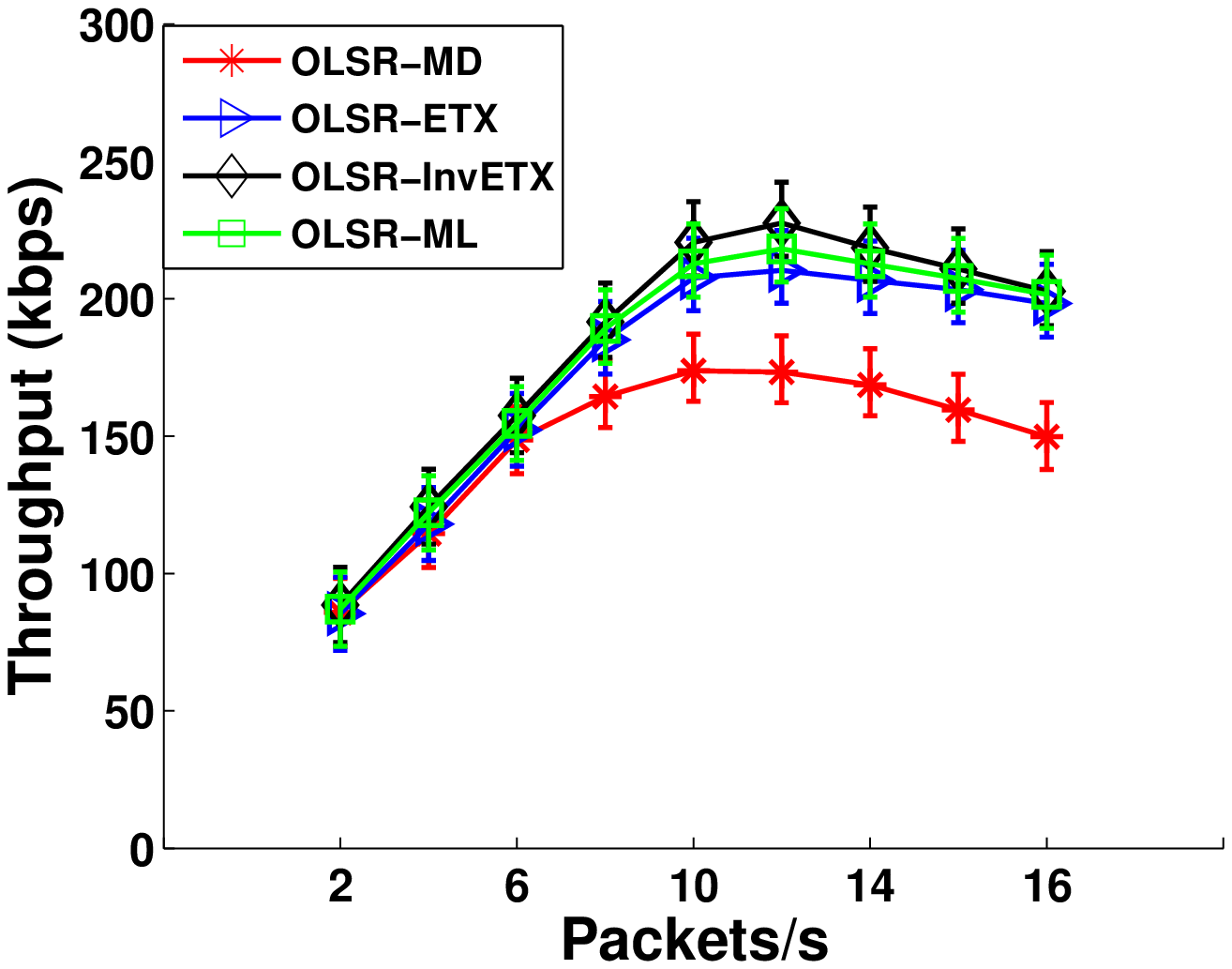}}
 \subfigure[Throughput of Enhanced OLSR with 4 metrics]{\includegraphics[height=4 cm,width=5 cm]{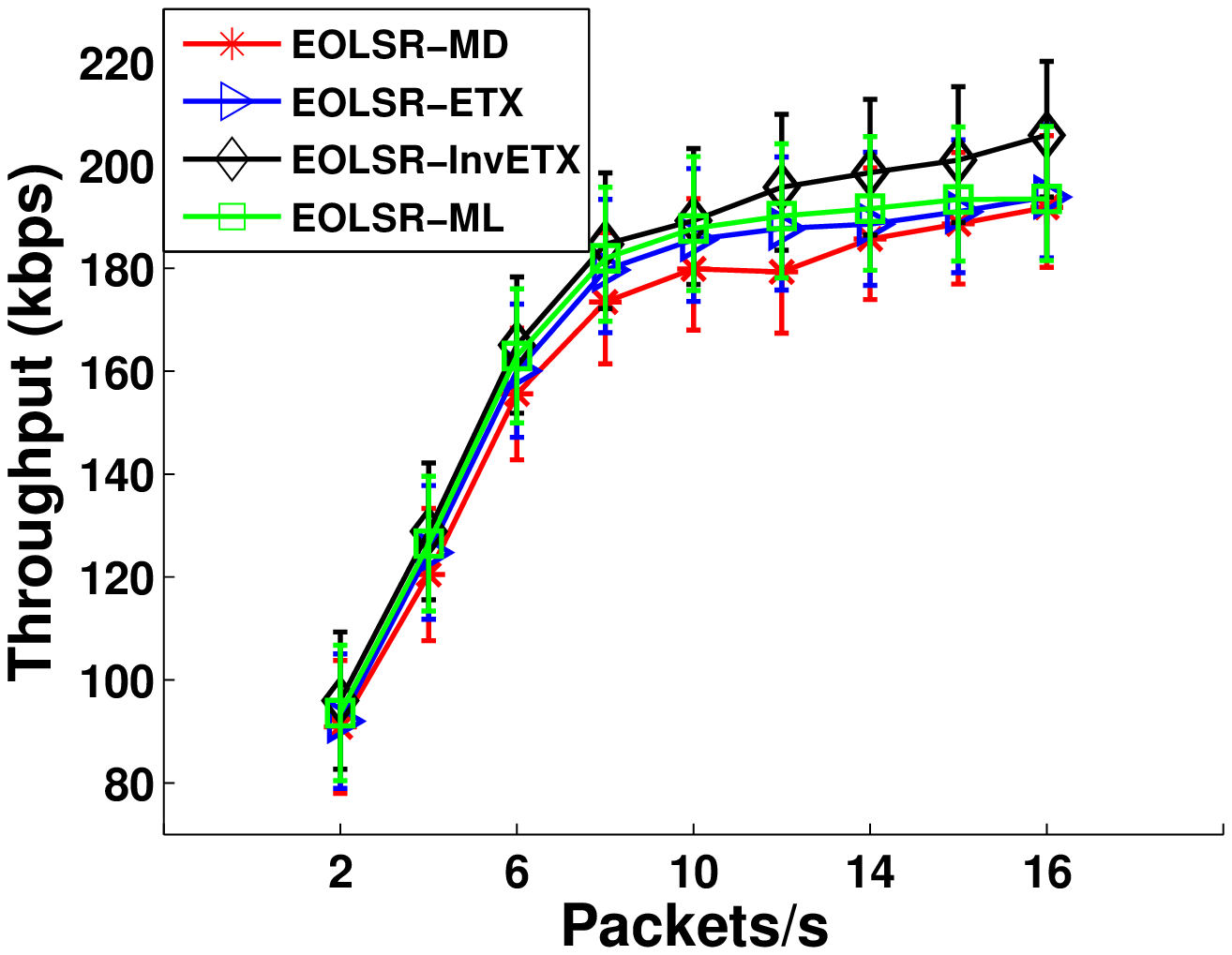}}
  \subfigure[E2ED of OLSR with 4 metrics]{\includegraphics[height=4 cm,width=5 cm]{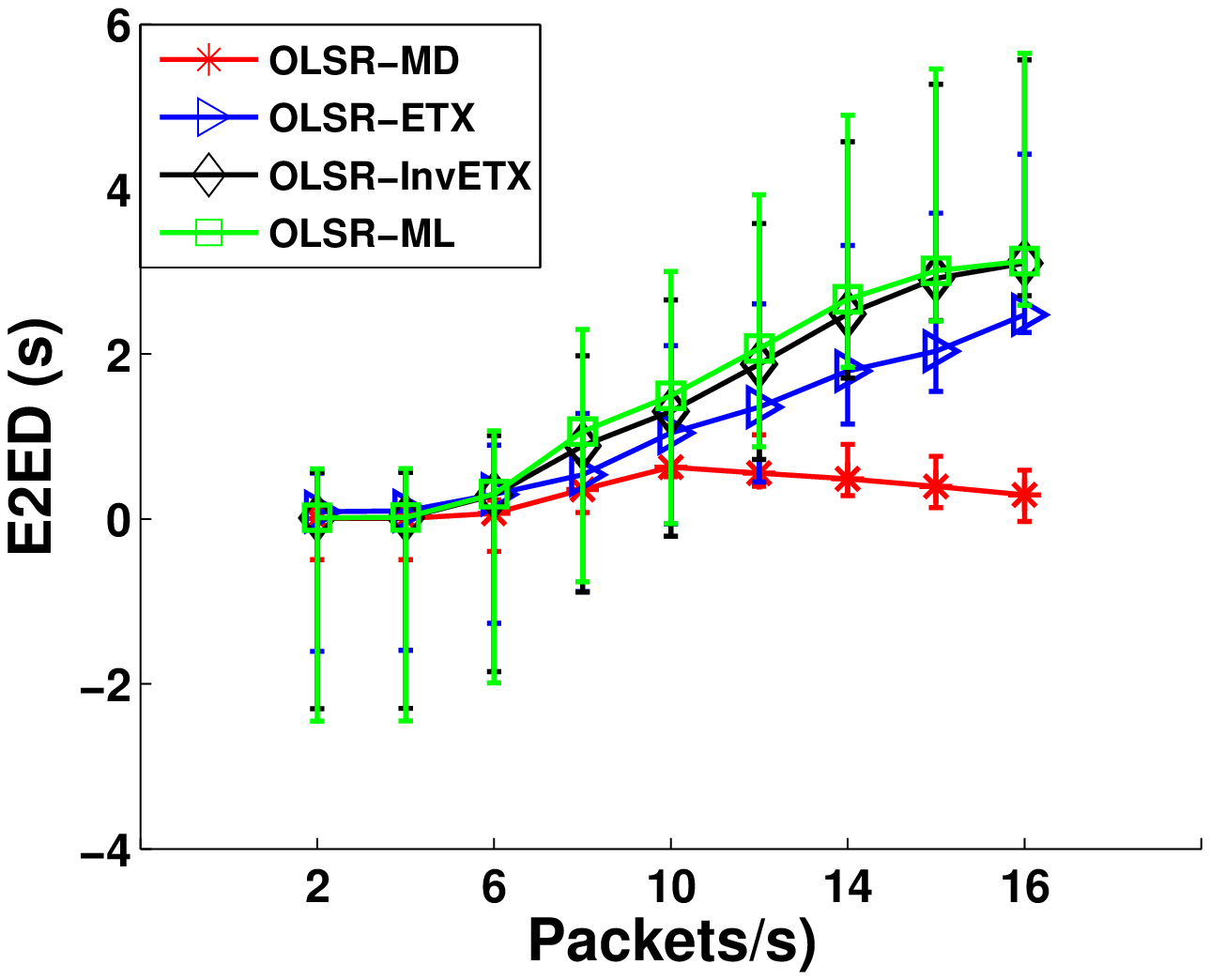}}
  \subfigure[E2ED of Enhanced OLSR with 4 metrics]{\includegraphics[height=4 cm,width=5 cm]{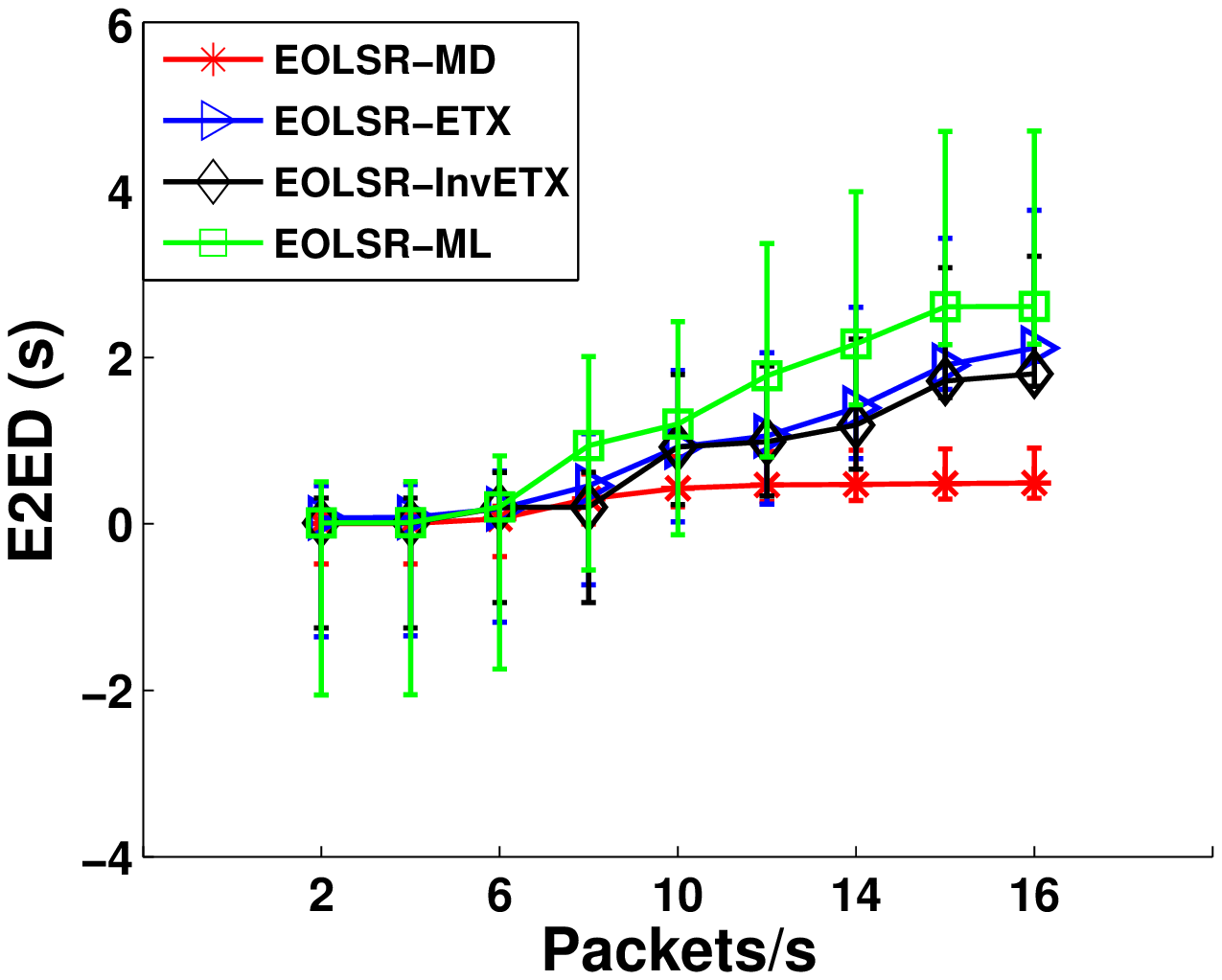}}
  \subfigure[NRL of OLSR with 4 metrics]{\includegraphics[height=4 cm,width=5 cm]{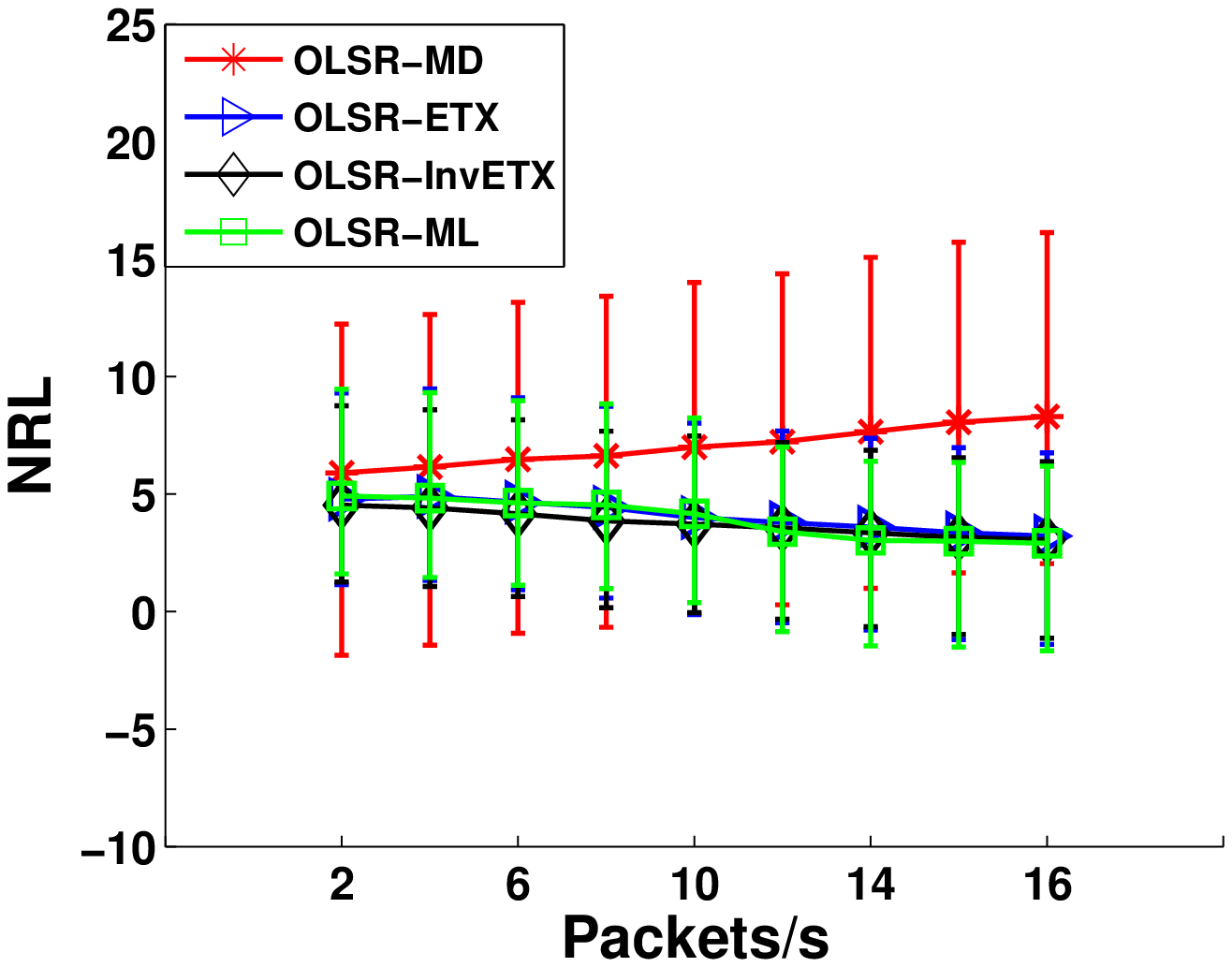}}
 \subfigure[NRL of Enhanced OLSR with 4 metrics]{\includegraphics[height=4 cm,width=5 cm]{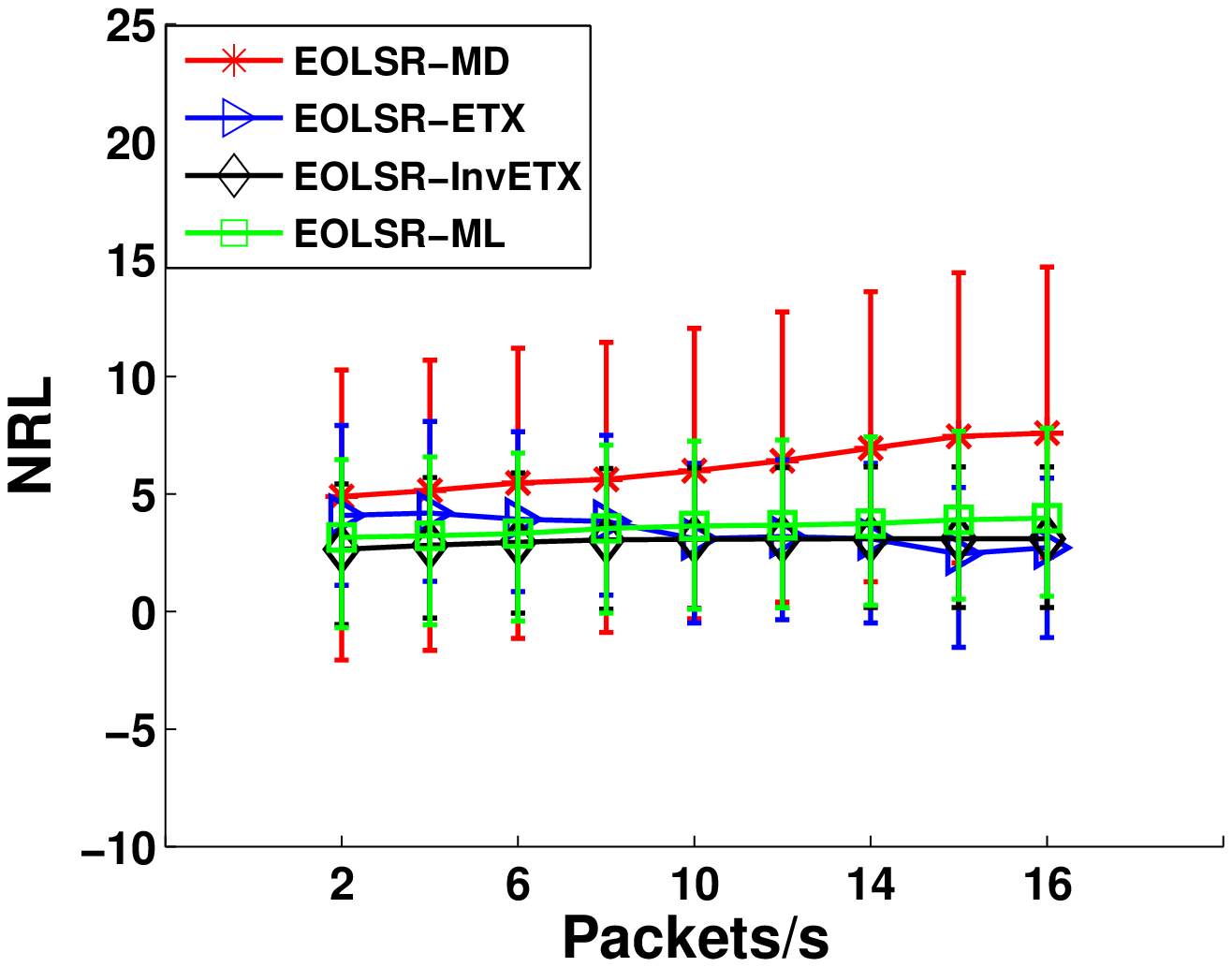}}
  \caption{Figure.1. Computational Overhead, Throughput, Delay, Routing load by Metrics}
\end{figure}

From [1], we analyze that routing load is a critical issue in SWMhNs when network load increases either with increase in number of nodes or with increase in number of packets. As, OLSR uses the shortest interval for exchanging topological information as compared to DSDV, i.e., 'full-dump-period' in DSDV, whereas, in OLSR $TC\_INTERVAL=5s$, therefore, it causes more routing overhead. In OLSR, after detecting any change in MPR's status, TC messages are triggered, thus, increase in $TC\_INTERVAL=15s$ value does not effect the stability of route.  By increasing $TC\_INTERVAL$'s, routing load is reduced. Trigger updates in OLSR depend on $HELLO\_INTERVAL$, therefore, to achieve stability in routing table, we decrease $HELLO\_INTERVAL$; $2s$ in OLSR, whereas, set to $1s$ in EOLSR. Another enhancement in EOLSR is that we set window $w=10$ instead of $20$.

\section{Simulation Results}
We use implementation of \textit{ETX}, \textit{MD} and \textit{ML} with default and enhanced version of OLSR in NS2-2.34. Then we implement the fourth metric, \textit{InvETX}, as expressed by eq. (2). In area of $1000m\times 1000$, $50$ nodes are placed randomly to form a static network. Constant Bit Rate (CBR) traffic is randomly generated by $20$ source-destination pairs with packet size of $64\,bytes$. Each simulation is run for five different topologies for $900\,s$ each. Then the average of five different values of each performance parameter is used to plot the graphs. To observe performance of OLSR with four metrics, we randomly generated the data traffic with number of packets from $2$ to $16$ per second.

To better understand the performance trade-offs, we take an example of the Static Wireless Multi-hop Networks (SWMhNs) that have two major issues; NRL and E2ED. In this type of networks, the proactive protocols are preferred due to periodically updation of network topology, like, OLSR, instead of the reactive ones. Moreover, hop-by-hop routing technique along with link state information help OLSR to handle aggressive routing overhead, as compared to source routing. Using MPRs selection along with proactive nature, OLSR reduces number of retransmissions and thus achieves minimum delay. In the following subsections, we discuss simulation results for conventional and enhanced OLSR (EOLSR) with respect to three performance parameters; throughput, E2ED, and NRL.

\subsection{Throughput}
In static networks, with varying data traffic rates, \textit{MD} produces lowest throughput, as compared to \textit{ETX}/\textit{InvETX} and \textit{ML}. Moreover, in medium and high network loads, there are more drop rates as compared to small load in the case of \textit{MD} metric. This is due to the one-way delays that are used to compute the \textit{MD} routing metric with small probe packets before setting up the routing topology and not considering the traffic characteristics. It may thus happen that, if no other traffic is present in the network, the probes sent on a link experience very small delays, but larger data packets may experience the higher delay or retransmission due to congestion. Thus, \textit{MD} is not suitable for the static networks with high traffic load, as, it degrades the network performance by achieving less throughput values. The \textit{ML} in medium and high network loads produces higher throughput values because \textit{ML} attains the less drop ratios, as compared to \textit{ETX}. Moreover, in \textit{ML} the paths with minimum loss rates or higher probabilities of successful (re)transmissions lead to high data delivery rates, with an additional advantage of more stable end-to-end paths and less drop rates.

\textit{MD} uses the Ad-hoc packet technique to measure the one-way delay. Then proactive delay assurance approach is used to measure \textit{MD} metric. The minimum delay metric performs best in terms of average packet loss probability. In Fig. 4(c), \textit{MD}'s delay is showing  the lowest values among other metrics. This is due to the route selection decision based on delay of ad-hoc probes. While \textit{ETX} and \textit{ML} produce increasing value of delay, when traffic increases. The very first reason is that both metrics have no mechanism to calculate the round trip, unlike \textit{MD} metric. Meanwhile, in \textit{ML}, selection of longer routes with high probability of successful transmission augments the delay, as compared to \textit{ETX}. In default OLSR, in high traffic rates more drop rates are noticed, as compared to low traffic rates. This is because of increase of routing overhead due to frequently generation of TC messages. In EOLSR, we change frequency of TC message generation by increasing value of $TC\_INTERVAL=15s$. This enhancement reduce congestion in the network and results in high throughput which is shown in Fig. 4(b) as compared to Fig. 4(a).

\subsection{E2ED}
The ad-hoc probe packets are sent by \textit{MD} to accurately measure the one-way delay. Thus, low latency is achieved by selecting the path with less Round Trip Time (RTT). On the other hand, these ad-hoc probes cause routing overhead in a network and decrease the throughput when data load is high in a static network. In static networks, to measure an accurate link with less routing load is a necessary condition. The delay cost due to increase in the number of intermediate hops is paid to achieve throughput by \textit{ML}. As \textit{ML} selects those paths which possess less loss rates, therefore, a longer path with high successful delivery is preferred. Thus the product of the link probabilities selection decreases the drop rates and increase the RTT.

\textit{ETX} uses the same mechanism to measure the link quality as that of \textit{ML}, i.e., modified HELLO messages. However, summing up the individual probabilities and preference of the shortest path reduces the delay of \textit{ETX} as compared to \textit{ML}. Thus, a slow link preference results more drop rates of \textit{ETX}, as compared to \textit{ML}. This sort of trade-off is common in routing protocols. While designing a link metric, if demands of the underlying network are taken into consideration then it becomes easy to decide that among which performance parameters, trade-off(s) should be made. For example, \textit{ML} and \textit{ETX} achieve higher throughput values than \textit{MD}, as shown in Fig. 1(b), whereas \textit{MD} remarkably achieves less E2ED than \textit{ML} and \textit{ETX} which is depicted in Fig. 4(c). In EOLSR, E2ED of the metrics becomes less as compared to default OLSR. In OLSR, HELLO messages are used to detect link status information. After detecting link failure, OLSR triggers TC message to update routing table. In default OLSR, $HELLO\_INTERVAL=2s$ is not enough to calculate recent information, and thus E2ED value is increased. On the other hand,  $HELLO\_INTERVAL=1s$ results in quick updation of routing table entries which consequently prevents path instabilities (Fig. 4(c), as compared to Fig. 4(d))

\subsection{NRL}
\textit{OLSR-MD} suffered from the highest routing loads. As, ad-hoc probes are used to measure the metric values and are sent periodically along with TC and HELLO messages. On the other hand, \textit{ETX} and \textit{ML} calculate the probabilities for the metric from the values obtained from the enhanced HELLO messages. OLSR uses HELLO and TC messages to calculate the routing table and these messages are sent periodically. The delivery ratios are measured using modified OLSR HELLO packets that are sent every $HELLO\_INTERVAL$.

Each node calculates the number of HELLO messages received in a $w$ second period ($w=20$, by default) and divides it by the number of HELLO messages that should have been received in the same period ($10$, by default). Each modified HELLO packet notifies the number of HELLO messages received by the neighbor during the last $w$ seconds, in order to allow each neighbor to calculate the reverse delivery ratio. The worse the link quality, the higher the \textit{ETX} link value. A link is perfect if the \textit{ETX} value is 1 and its packet delivery fraction is also $1$, i.e., no packet loss. On the other hand, if in $w seconds$ period a node has not received any HELLO message then \textit{ETX} is set to $0$ and the link is not considered for routing due to 100\% loss ratio. Thus, due to no extra overhead to measure the metric \textit{ETX}/\textit{invETX} and \textit{ML} have to suffer from low routing load as compared to \textit{MD}. In EOLSR, for path stabilities we have changed $HELLO\_INTERVAL=1s$ and value of $w=10$. This enhancement results in quick updation, however, we also reduce frequency of $TC\_INTERVAL$ to reduce routing overhead which is produced by frequent emission of $HELLO$ messages. Thus, overall path stabilities results low routing overhead which can be seen in Fig. 1(e) as compared to Fig. 1(f).

\section{Conclusion}
In this work we select four quality link metrics; Expected Transmission Count (ETX), Minimum Delay (MD), Minimum Loss (ML) and our proposed meric; Inverse ETX (InvETX) with OLSR. We discuss several possible issues regarding wireless networks that can better help in designing a link metric. The ambition of a high throughput network can only be achieved by targeting a concrete compatibility of the underlying wireless network, the routing protocol operating it, and routing metric; heart of a routing protocol. Depending upon the most demanding features of the networks, different routing protocols impose different costs of 'message overhead' and 'management complexity'. These costs help to understand that which type of routing protocol is well suitable for which kind of underlying wireless network and then which routing link metric is appropriate for which routing protocol. A novel contribution of this paper is enhancement in original OLSR; EOLSR, to achieve high efficiency in terms of optimum routing load and routing latency. For this purpose, first we present a mathematical framework, and then to validate this frame work, selected metrics are simulated with OLSR and EOLSR. For comparison three important performance parameters are selected; throughput, NRL and E2ED. From our simulation results, we conclude that by adjusting the frequencies of exchanging topological information high efficiency can be achieved.


\begin{thebibliography}{00}
\bibitem{1} Javaid, N.; Ullah, M.; Djouani, K.;, ``Identifying Design Requirements for Wireless Routing Link Metrics'', Global Telecommunications Conference (GLOBECOM 2011), 2011 IEEE , vol., no., pp.1-5, 5-9 Dec. 2011.
\bibitem{2} D. S. J. de Couto, ``High-throughput routing for multi-hop wireless networks'', Ph.D. dissertation, MIT, 2004.
\bibitem{3}Moreira, W., Aguiar, E., Abelém, A., Stanton, M., ``Using multiple metrics with the optimized link state routing protocol for wireless mesh networks'', Simpósio Brasileiro de Redes de Computadorese Sistemas Distribuídos, Maio (2008).
\bibitem{4}D. Passos, D. V. Teixeira, D. C. Muchaluat-Saade, L. C. S. Magalhaes, and C. V. N. de Albuquerque, ``Mesh network performance measurements'', I2TS, 2006.

\bibitem{5}	Clausen, Thomas, Philippe Jacquet, Cédric Adjih, Anis Laouiti, Pascale Minet, Paul Muhlethaler, Amir Qayyum, and Laurent Viennot. "Optimized link state routing protocol (OLSR)." (2003).


\bibitem{6} Javaid, Nadeem. ``Analysis and design of quality link metrics for routing protocols in Wireless Networks.'' PhD diss., Université Paris-Est, 2010.

\bibitem{7} Clausen, T., and P. Jaqcquet. ``Optimized link state routing (OLSR) RFC 3626.'' IETF Networking Group (October 2003) (2003).

\bibitem{8} Javaid, Nadeem, Ayesha Bibi, and Karim Djouani. ``Interference and bandwidth adjusted ETX in wireless multi-hop networks.'' In GLOBECOM Workshops (GC Wkshps), 2010 IEEE, pp. 1638-1643. IEEE, 2010.

\bibitem{9} C. E. Perkins and P. Bhagwat, ``Highly dynamic Destination-Sequenced Distance-Vector routing (DSDV) for mobile computers'', SIGCOMM Comput. Commun. Rev., vol. 24, pp. 234-244, 1994.

\bibitem{10} Yaling Yang, Jun Wang, and Robin Kravets, ``Interference-aware Load Balancing for Multihop Wireless Networks'', Tech. Rep. UIUCDCS-R-2005-2526, Department of Computer Science, University of Illinois at Urbana-Champaign, 2005.

\bibitem{11} Das, Saumitra M., Yunnan Wu, Ranveer Chandra, and Y. Charlie Hu. "Context-based Routing: Technique, Applications, and Experience." In NSDI, vol. 8, pp. 379-392. 2008.

\bibitem{12} R. Draves, J. Padhye, and B. Zill, ``Routing in multi-radio, multi-hop wireless mesh networks'', in ACM International Conference on Mobile Computing and Networking (MobiCom), Sept. 2004, pp. 114-128.

\bibitem{13} Richard Draves, Jitendra Padhye, Brian Zill, ``Comparison of routing metrics for static multi-hop wireless networks'', Vol. 34, No. 4. (October 2004), pp. 133-144.

\bibitem{14} Yang, Yaling, Jun Wang, and Robin Kravets. "Designing routing metrics for mesh networks." In IEEE Workshop on Wireless Mesh Networks (WiMesh). 2005.

\bibitem{15} Javaid, Nadeem, Ayesha Bibi, Akmal Javaid, and Shahzad A. Malik. "Modeling routing overhead generated by wireless reactive routing protocols." In Communications (APCC), 2011 17th Asia-Pacific Conference on, pp. 631-636. IEEE, 2011.

\bibitem{16} Sagar, S., J. Saqib, A. Bibi, and N. Javaid. "Evaluating and Comparing the Performance of DYMO and OLSR in MANETs and in VANETs." In Multi-topic Conference (INMIC), 2011 IEEE 14th International, pp. 362-366. IEEE, 2011.

\bibitem{17} Wasiq, S., W. Arshad, N. Javaid, and A. Bibi. "Performance evaluation of DSDV, OLSR and DYMO using 802.11 and 802. lip MAC-protocols." In Multitopic Conference (INMIC), 2011 IEEE 14th International, pp. 357-361. IEEE, 2011.

\bibitem{18} Javaid, Nadeem, M. Yousaf, A. Ahmad, A. Naveed, and Karim Djouani. "Evaluating impact of mobility on wireless routing protocols." In Wireless Technology and Applications (ISWTA), 2011 IEEE Symposium on, pp. 84-89. IEEE, 2011.

\bibitem{19} Javaid, N.; Bibi, A.; Javaid, A.; Malik, S.A., "Modeling routing overhead generated by wireless proactive routing protocols," GLOBECOM Workshops (GC Wkshps), 2011 IEEE , vol., no., pp.1072,1076, 5-9 Dec. 2011. doi: 10.1109/GLOCOMW.2011.6162343.

%
%
\end{thebibliography}
\end{document}